\documentclass{pasj01}
\draft 
\Received{$\langle$reception date$\rangle$}
\Accepted{$\langle$acception date$\rangle$}
\Published{$\langle$publication date$\rangle$}
\begin{document}

\title{Sub-millimeter Detection of a Galactic Center Cool Star IRS 7 by ALMA}
\author{Masato Tsuboi$^{1, 2}$, Yoshimi Kitamura$^1$,  Takahiro Tsutsumi$^3$, Ryosuke Miyawaki$^4$, Makoto Miyoshi$^5$ and Atsushi Miyazaki$^6$ }%
\altaffiltext{1}{Institute of Space and Astronautical Science, Japan Aerospace Exploration Agency,\\
3-1-1 Yoshinodai, Chuo-ku, Sagamihara, Kanagawa 252-5210, Japan }
\email{tsuboi@vsop.isas.jaxa.jp}
\altaffiltext{2}{Department of Astronomy, The University of Tokyo, Bunkyo, Tokyo 113-0033, Japan}
\altaffiltext{3}{National Radio Astronomy Observatory,  Socorro, NM 87801-0387, USA}
\altaffiltext{4}{College of Arts and Sciences, J.F. Oberlin University, Machida, Tokyo 194-0294, Japan}
\altaffiltext{5}{National Astronomical Observatory of Japan, Mitaka, Tokyo 181-8588, Japan}
\altaffiltext{6}{Japan Space Forum, Kanda-surugadai, Chiyoda-ku,Tokyo,101-0062, Japan}
\maketitle
\begin{abstract}
IRS 7 is an M red supergiant star which is located at $5\farcs5$ north of Sagittarius A$^\ast$. We detected firstly the continuum emission at 340 GHz of IRS 7 using ALMA. The total flux density of IRS 7 is $S_\nu=448\pm45 \mu$Jy. The flux density indicates that IRS  7 has a photosphere radius of $R=1170\pm60 ~R_\odot$, which is roughly consistent with the previous VLTI measurement. 
We also detected a shell like feature with north extension in the H30$\alpha$ recombination line by ALMA. 
The electron temperature and electron density of the shell like structure are estimated to be $\bar{T}^\ast_{\mathrm e}=4650\pm500$ K and  $\bar{n}_{\mathrm e}=(6.1\pm0.6)\times10^4$ cm$^{-3}$, respectively. The mass loss rate is estimated to be $\dot{m} \sim 1\times 10^{-4} M_\odot$ yr$^{-1}$,  which is  consistent with a typical mass loss rate of a pulsating red supergiant star with $M=20-25 M_\odot$.
The kinematics of the ionized gas would support the hypothesis that the shell like structure made by the mass loss of IRS 7 is supersonically traveling in the ambient matter toward the south.  
The brightened southern half of the structure and the north extension would be a bow shock and a cometary-like tail structure, respectively. 
\end{abstract}

\section{Introduction}
The Galactic center region is the nucleus of the nearest spiral galaxy and harbors the Galactic center black hole, Sagittarius A$^\ast$ (Sgr A$^\ast$),  with $M\sim4\times10^6 M_\odot$ (e.g. \cite{Ghez}, \cite{Boehle}, \cite{Abuter}). 
This region is recognized to be a laboratory for peculiar phenomena, which will be found in the central regions of external spiral galaxies by future telescopes.
A  bright star cluster within $r<0.5$ pc of Sgr A$^\ast$ has been found by early IR observations (e.g. \cite{Genzel},  \cite{Figer1999},  \cite{Figer2002}), which is called the Nuclear star cluster (NSC).  The NSC contains over several tens Wolf-Rayet and OB stars.  IRS 7 is a bright IR star which is located at $5\arcsec.5$ north of Sgr A$^\ast$. Although IRS 7 has been thought to be a member star of the NSC, this is classified into an M1(e.g. \cite{Paumard2014}) or M2(e.g. \cite {Carr}, \cite{Perger}) red supergiant star based on IR spectroscopic observations. The SiO maser emission is often detected toward such late-type stars in the Galactic disk region and the maser emission also has been detected toward IRS 7 (e.g. \cite{Reid2003}, \cite{Borkar}). 
In radio wavelength, IRS 7 has been observed as a bow-shock like feature with north extension (e.g. \cite{Yusef-Zadeh1992}, \cite{Yusef-Zadeh2015}).   This feature is thought to be a cometary-like structure made  by strong UV radiation from  early-type stars in the NSC. However, IRS 7 itself has not yet has been detected in radio wavelength although many early-type stars in the NSC have been detected  (e.g. \cite{Yusef-Zadeh2014}, \cite{Moser}). 

The continuum emission from the photosphere of IRS 7 should increase with increasing observation frequency because this emission is optically thick thermal one. 
Because the line to continuum ratio of Hydrogen recombination line should increase with increasing observation frequency in radio observations of millimeter or sub-millimeter wavelengths such observations are  suitable to detect both the continuum emission from IRS 7 itself and the recombination line from the extended structure.
ALMA is the most powerful tool to observe these in the wavelengths. Therefore we have observed IRS 7 at 340 GHz using ALMA and analyzed the data set of the Sgr A$^\ast$ region obtained in the Director's Discretionary Time (DDT) observation of ALMA, which involves the H30$\alpha$ recombination line.  
Throughout this paper, we adopt $d\sim8.2$ kpc as the distance to the Galactic center (e.g. \cite{Boehle}, \cite{Abuter}). Then, $1\arcsec$ corresponds to about 0.04 pc at the distance.

\section{Observation and Data Reduction}
\subsection{340 GHz  Observation}
We have observed Sgr A$^\ast$ itself and the surrounding area including the NSC  at 340 GHz as an ALMA Cy.3 program (2015.1.01080.S. PI M.Tsuboi). The data has been published already in the previous paper, specifically for  the study of IRS 13E (\cite{Tsuboi2017b}). Therefore the observations are summarised briefly here.
The FOV is centered at  $\alpha_{\rm ICRS}$ = $17^{\rm h}45^{\rm m}40^{\rm s}.04$  and $\delta_{\rm ICRS}$= $-29^{\circ}00'28\farcs2$, which is a nominal center position of  Sgr A$^\ast$. 
The diameter of the FOV is  $\sim18\arcsec$ at 340 GHz  in FWHM. 
The observations were performed in three days (23 Apr. 2016, 30/31 Aug. 2016, and 08 Sep. 2016). The observation in April was for detection of extended emission. 
 J1717-3342 was used as a phase calibrator in the 340 GHz observation.  The flux density scale was determined using Titan and J1733-1304. 
We performed the data analysis by Common Astronomy Software Applications (CASA) \citep{McMullin}. The complex gain errors of the data were minimized using the ``self-calibration" method in CASA 5.4. 
The imaging to obtain the maps was done using CASA 5.4 with {\tt clean} task. 
Using two days data  of Aug. and  Sep. 2016, the synthesized beam size(angular resolution) using ``briggs weighting (robust parameter = 0.2)"  is $\theta_\mathrm{maj} \times \theta_\mathrm{min}=0\farcs107 \times 0\farcs101, PA=-78^\circ$ in FWHM. 
The $1\sigma$ sensitivity in the emission-free area is  $45~\mu$Jy beam$^{-1}$.  This sensitivity corresponds to 0.04 K in $T_\mathrm{B}$. 
The dynamic range reaches to  $>50000$ in the area.

\subsection{H30$\alpha$ recombination line}
We analyzed the DDT observation of the Sgr A$^\ast$ region with ALMA (2015.A.00021.S). The data set involves the H30$\alpha$ recombination line ($\nu_{rest}$= 231.9009 GHz).  The analysis also has been published already in the previous paper, specifically for  the study of gas motions around IRS 13E (\cite{Tsuboi2017b}). Therefore the observations are summarised briefly here.
The FOV is 25$\arcsec$, which is centered at Sgr A$^\ast$. J1744-3116 was used as a phase calibrator. The flux density scale was determined using Titan and J1733-1304. After the subtraction of the continuum emission from the data using CASA task, {\tt mstransform (fitorder=1)}, we made images using CASA 5.0 with {\tt tclean} task with the multi-threshold auto-boxing algorithm ( {\tt usemask=`auto-multithresh'} ) which is automatically identify the emission regions to be CLEANed using threshold based on rms noise and sidelobe level and updated as the deconvolution iterations progress \citep{Kepley}.  Using ``briggs weighting (robust parameter = 0.5)",  we obtained an H30$\alpha$ recombination line data cube with small synthesized beam size (angular resolution) ($\theta_\mathrm{maj} \times \theta_\mathrm{min}=0\farcs41 \times 0\farcs30, PA = -77^{\circ}$ in FWHM) and high sensitivity (0.2 mJy beam$^{-1}$ at a line-free channel).  The sensitivity is close to the expected one by the ALMA sensitivity calculator. 

\begin{figure}
\begin{center}
\includegraphics[width=18cm, bb=0 0 532.1 464.11 ]{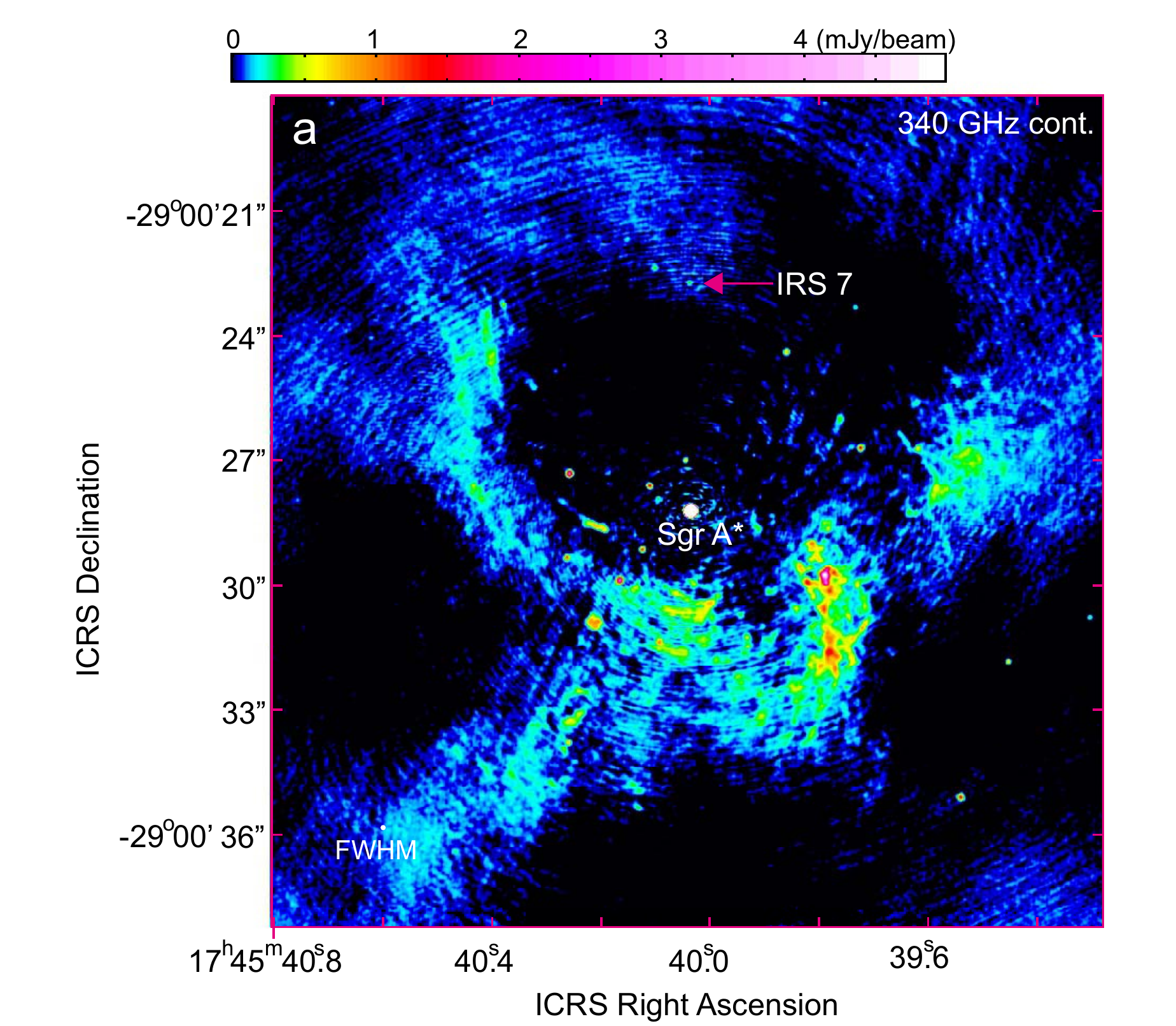}
\end{center}
\caption{{\bf a} Continuum map at 340 GHz of the Sgr A$^\ast$ region.   Using 2D Gaussian fitting, the position of Sgr A$^\ast$ is derived to be $\alpha_{\rm ICRS}=17^{\rm h}45^{\rm m}40^{\rm s}.03389\pm0^{\rm s}.00003$,  $\delta_{\rm ICRS }=-29^\circ00\arcmin28\farcs2243\pm0\farcs0003$  at the observation epoch of $2016.68$.  This is a finding chart of IRS 7.   
The angular resolution is $0\farcs107 \times 0\farcs101, PA=-78^\circ$  in FWHM, which is shown as the oval at the lower left corner.   The $1\sigma$ noise level is $0.045$ mJy beam$^{-1}$. }
\end{figure}
\begin{figure}
\addtocounter{figure}{-1}
\begin{center}
\includegraphics[width=18cm, bb=0 0 532.11 452.93 ]{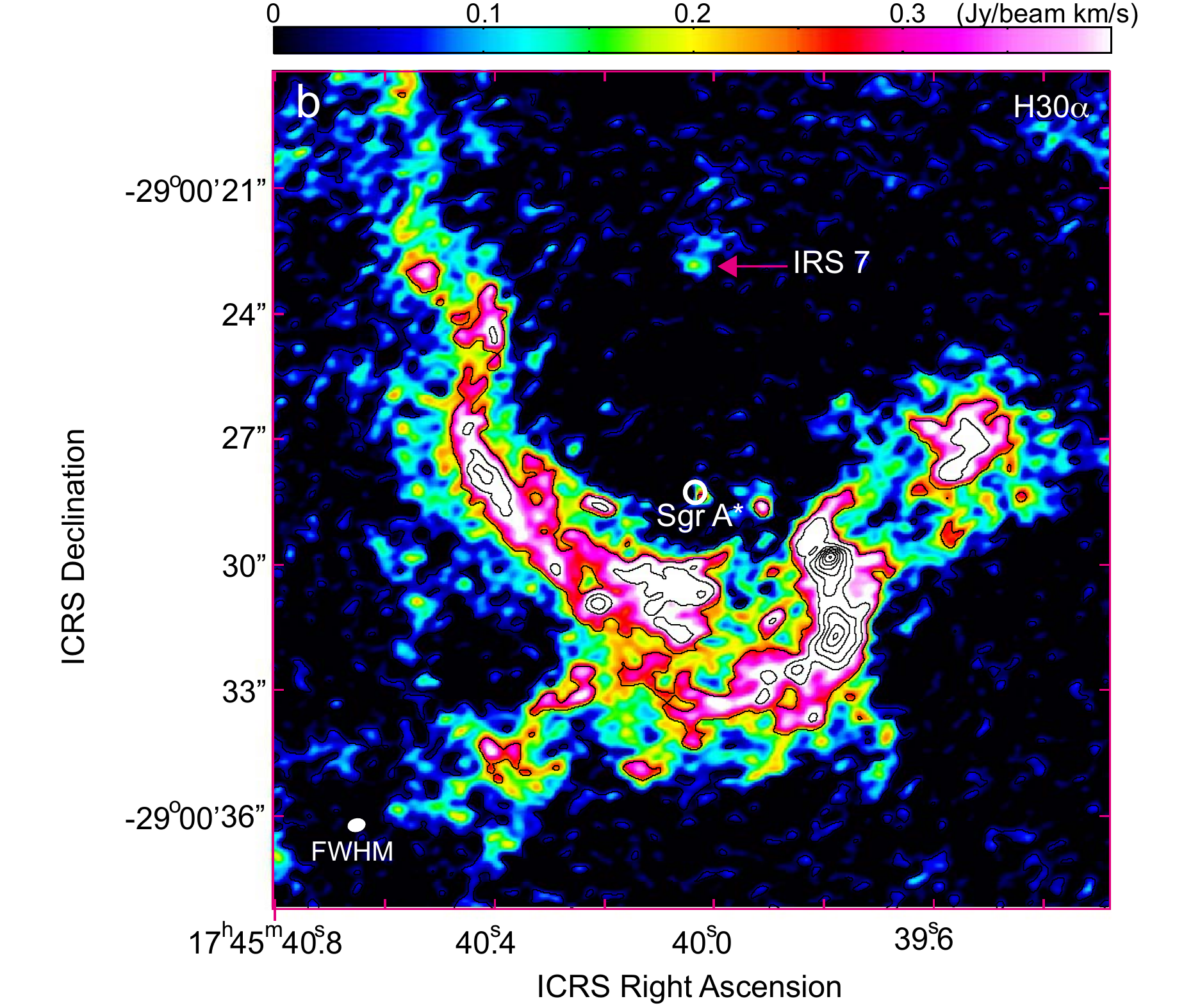}
\end{center}
\caption{{\bf b} Moment 0 map of the Sgr A$^\ast$ region in the H30$\alpha$ recombination line. The  integrated velocity range   is $V_\mathrm{LSR} = -400$ to $400$ km s$^{-1}$. The angular resolution is $0\farcs41 \times 0\farcs30, PA=-77^\circ$  in FWHM, which is shown as the oval at the lower left corner.  The white circle shows the position of Sgr A$^\ast$.  The $1\sigma$ noise level is $0.05$ Jy beam$^{-1}$ km s$^{-1}$. }
\end{figure}

\begin{figure}
\begin{center}
\includegraphics[width=13cm, bb=0 0 594.05 428.17]{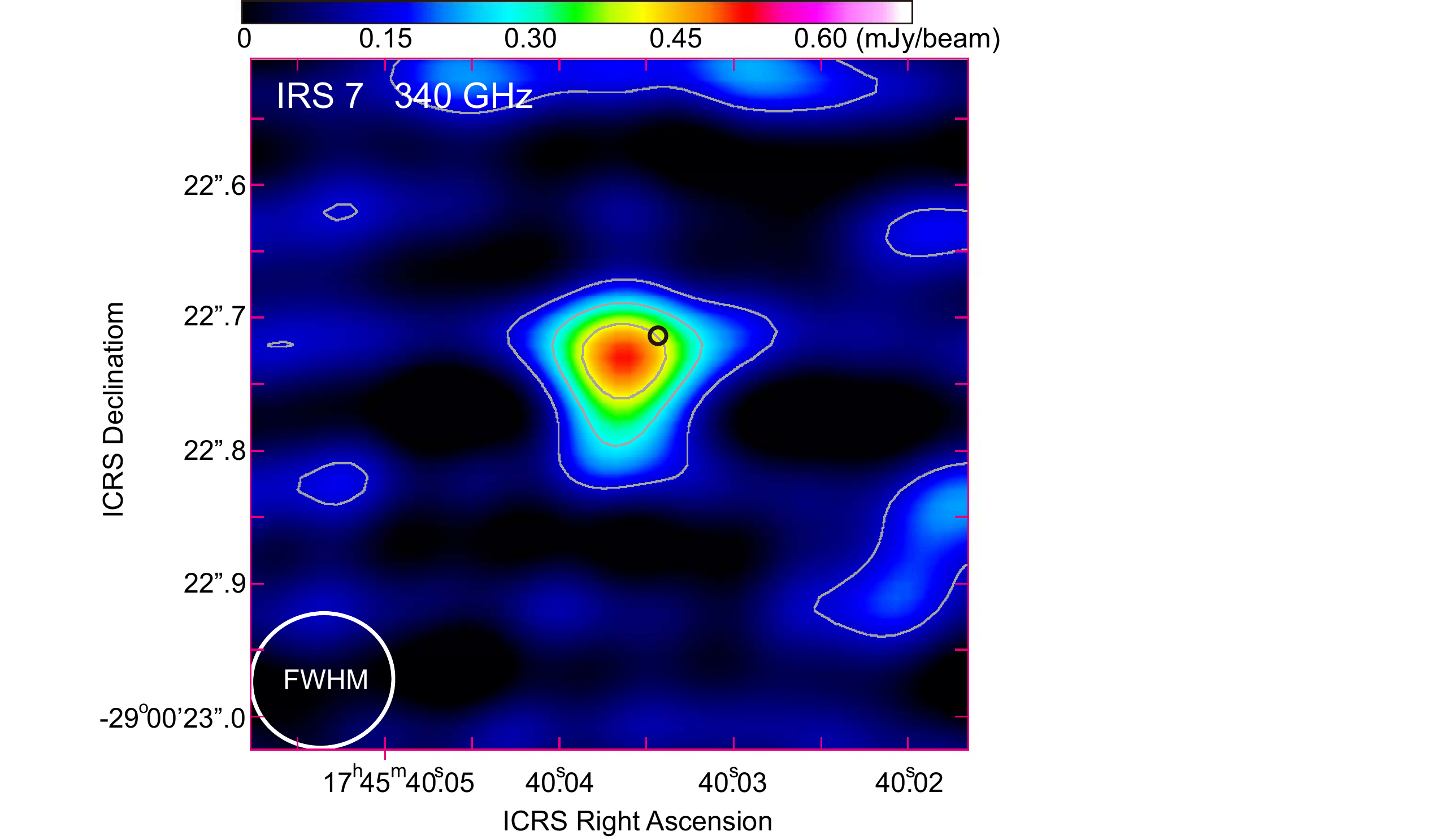}
\end{center}
\caption{Close-up continuum map of IRS 7 at 340 GHz.  The angular resolution is  $0\farcs107 \times 0\farcs101, PA=-78^\circ$ in FWHM, which is shown as the white oval at the lower left corner.   The $1\sigma$ noise level is $0.045$ mJy beam$^{-1}$. The lowest color level corresponds to $3\sigma$.  The position of IRS 7 is $\alpha_{\rm ICRS}=17^{\rm h}45^{\rm m}40^{\rm s}.03625\pm0^{\rm s}.0002$,  $\delta_{\rm ICRS}=-29^\circ00\arcmin22\farcs732\pm0\farcs002$  at the observation epoch of $2016.68$. The peak intensity is $I_\nu=448\pm45 \mu$Jy beam$^{-1}$ after primary-beam correction. This value corresponds to  $T_\mathrm{B}=0.41\pm0.04$ K.  The black circle  indicates the  IR position  at the observation epoch, which is corrected to be $\alpha_{\rm ICRS}=17^{\rm h}45^{\rm m}40^{\rm s}.034$,  $\delta_{\rm ICRS}=-29^\circ00\arcmin22\farcs71$ using the IR proper motion (\cite{Schodel}). }
\end{figure}
 
\section{Results}  
Figure 1a shows the continuum map at 340 GHz of the Sgr A$^\ast$ region. The angular resolution is $\theta_{{\mathrm maj}}\times \theta_{{\mathrm min}}=0\farcs107 \times 0\farcs101, PA=-78^\circ$ in FWHM.  Using 2D Gaussian fitting, the position of Sgr A$^\ast$ is derived to be $\alpha_{\rm ICRS}=17^{\rm h}45^{\rm m}40^{\rm s}.03389\pm0^{\rm s}.00003$,  $\delta_{\rm ICRS }=-29^\circ00\arcmin28\farcs2243\pm0\farcs0003$  at the observation epoch of $2016.68$. This figure is a finding chart of IRS 7.   Figure 2 shows the close-up continuum map of IRS 7 at 340 GHz.  The angular resolution is  the same as that  in Figure 1a. The continuum emission of IRS 7 itself was identified as a point-like source in the map although the image is slightly distorted by unavoidable side lobes of Sgr A$^\ast$.
This is the first sub-millimeter detection of a cool star in the NSC. Using 2D Gaussian fitting, the position of IRS 7 is derived to be $\alpha_{\rm ICRS}=17^{\rm h}45^{\rm m}40^{\rm s}.03625\pm0^{\rm s}.0002$,  $\delta_{\rm ICRS }=-29^\circ00\arcmin22\farcs732\pm0\farcs002$  at the observation epoch of $2016.68$. 
The position is consistent with that by IR observations  (e.g. \cite{Genzel},  \cite{Figer1999}, \cite{Figer2002}, \cite{Schodel}).  
 The black circle in the figure indicates the IR position at the observation epoch, which is corrected to be $\alpha_{\rm ICRS}=17^{\rm h}45^{\rm m}40^{\rm s}.034$,  $\delta_{\rm ICRS}=-29^\circ00\arcmin22\farcs71$ using the IR proper motion (\cite{Schodel}). The angular extent of IRS 7 itself is not detected, and thus the total flux density  at 340 GHz of IRS 7 is estimated to be $S_\nu=448\pm45 \mu$Jy  after primary-beam correction.  This flux density corresponds to  the  beam averaged brightness temperature of $\bar{T_\mathrm{B}}=0.41\pm0.04$ K using the following formula;
\begin{equation}
\label{ }
\bar{T_\mathrm{B}}[{\mathrm K}]=1.22\times10^6\Big(\frac{\theta_{{\mathrm maj}}\times \theta_{{\mathrm min}}}{\mathrm{arcsec}^2} \Big)^{-1}\Big(\frac{\nu}{\mathrm{GHz}}\Big)^{-2}S_\nu[{\mathrm{ Jy }}]
\end{equation} 

Figure 1b shows the moment 0 map in the H30$\alpha$ recombination line. The integrated velocity range  is $V_\mathrm{LSR} = -400$ to $400$ km s$^{-1}$. The angular resolution is $\theta_{{\mathrm maj}}\times \theta_{{\mathrm min}}=0\farcs41 \times 0\farcs30, PA=-77^\circ$  in FWHM.  The white circle in the map shows the position of Sgr A$^\ast$. An  ionized gas structure  surrounding  IRS 7 is extended toward north. This ionized gas structure is consistent with the extended continuum emission feature detected in radio wavelength (e.g. \cite{Yusef-Zadeh1992},  \cite{Yusef-Zadeh2015}).

Figure 3 shows the close-up channel maps of IRS 7 in the H30$\alpha$ recombination line (pseudocolor).  
The $1\sigma$ noise level is $0.2$ mJy beam$^{-1}$. The lowest color level corresponds to $3\sigma$.
These are the first spatially and velocity resolved images of the ionized gas of IRS 7 (Cf. \cite{Zhao2010} ).
The contours show the continuum emission at 340 GHz for comparison. The velocity width of each map  is $\Delta V=10$ km s$^{-1}$. The central velocity ranges from $V_\mathrm{c, LSR} = -90$ to $-200$ km s$^{-1}$. The angular resolution is  the same as that  in Figure 1b. 
In the H30$\alpha$ recombination line, the extended structure surrounding IRS 7 is identified as a shell like structure with north extension. The shell like structure is seen in the panels with the central velocities of $V_\mathrm{c, LSR} = -100$ to $-150$ km s$^{-1}$. The diameter of the structure is $\theta_\mathrm{d}=1\farcs38$ after synthesized beam subtraction or $D=0.055$ pc. The southern half of the structure is brighter than the northern half.
While the north extension is seen in the panels with the central velocities of $V_\mathrm{c, LSR} = -140$ to $-190$ km s$^{-1}$. The position of  the north extension shifts from the vicinity of the shell like structure to north with decreasing  velocity. The structure over $3\sigma$ is extended up to $\Delta\theta=3\farcs3$ from IRS 7 in projection, which corresponds to $0.13$  pc. 

\begin{figure}
\begin{center}
\includegraphics[width=15cm, bb=0 0 1081.04 1469.29]{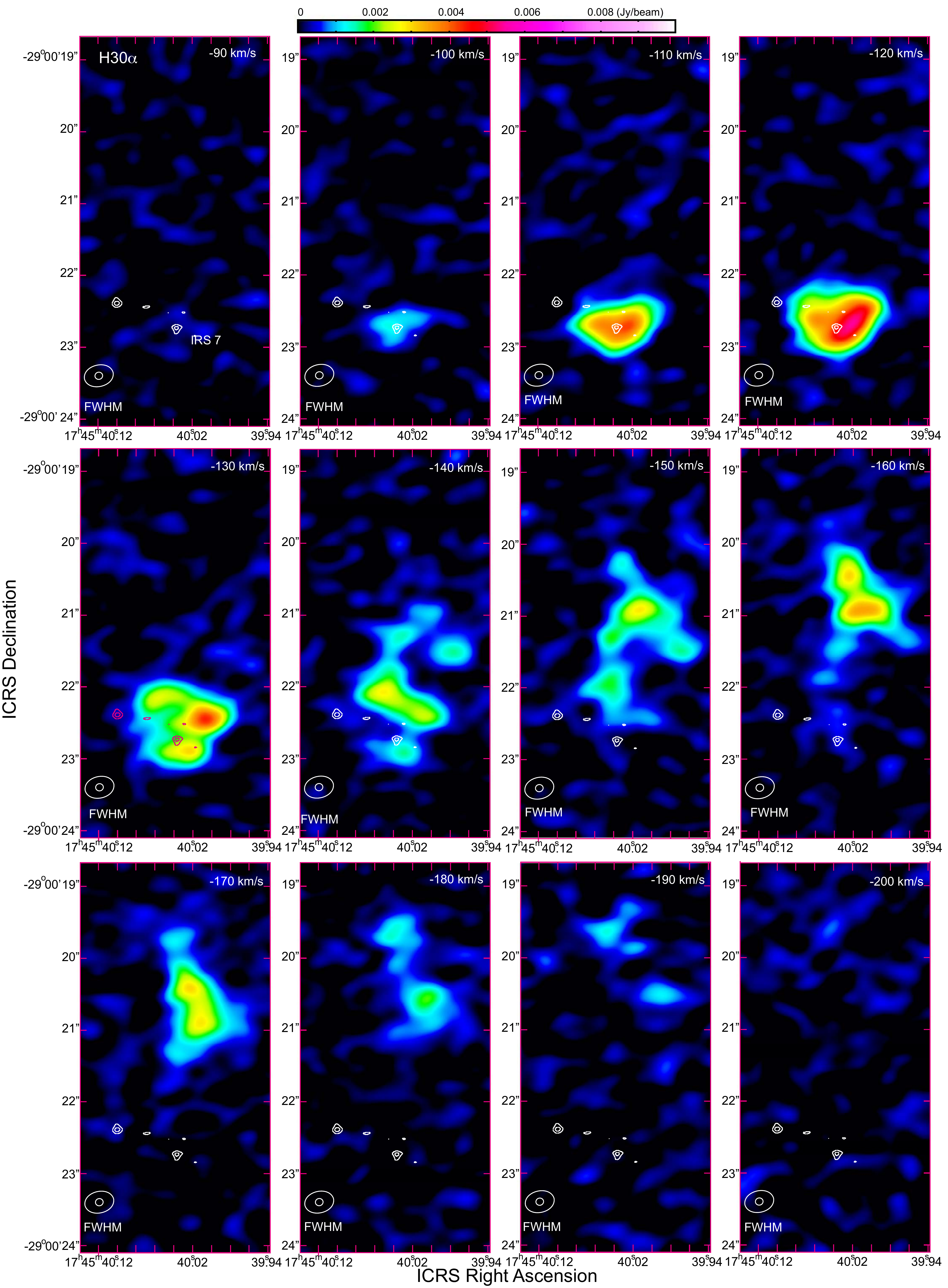}
\end{center}
\caption{Close-up channel map of IRS 7 in the H30$\alpha$ recombination line (pseudocolor).  The velocity range   is $V_\mathrm{LSR} = -90$ to $-200$ km s$^{-1}$. The velocity width of each maps  is $\Delta V=10$ km s$^{-1}$. The angular resolution is $0\arcsec.41 \times 0\arcsec.30, PA=-77^\circ$  in the H30$\alpha$ recombination line,  which is shown as the  large oval at the lower left corner.    The $1\sigma$ noise level is $0.2$ mJy beam$^{-1}$. The lowest color level corresponds to $3\sigma$.  The contours show the continuum emission at 340 GHz for comparison.  The angular resolution of the 340 GHz continuum emission is $0\farcs107 \times 0\farcs101, PA=-78^\circ$, which is shown as the small oval at the lower left corner.} 
\end{figure}

\begin{figure}
\begin{center}
\includegraphics[width=15cm, bb=0 0 753.21 832.18]{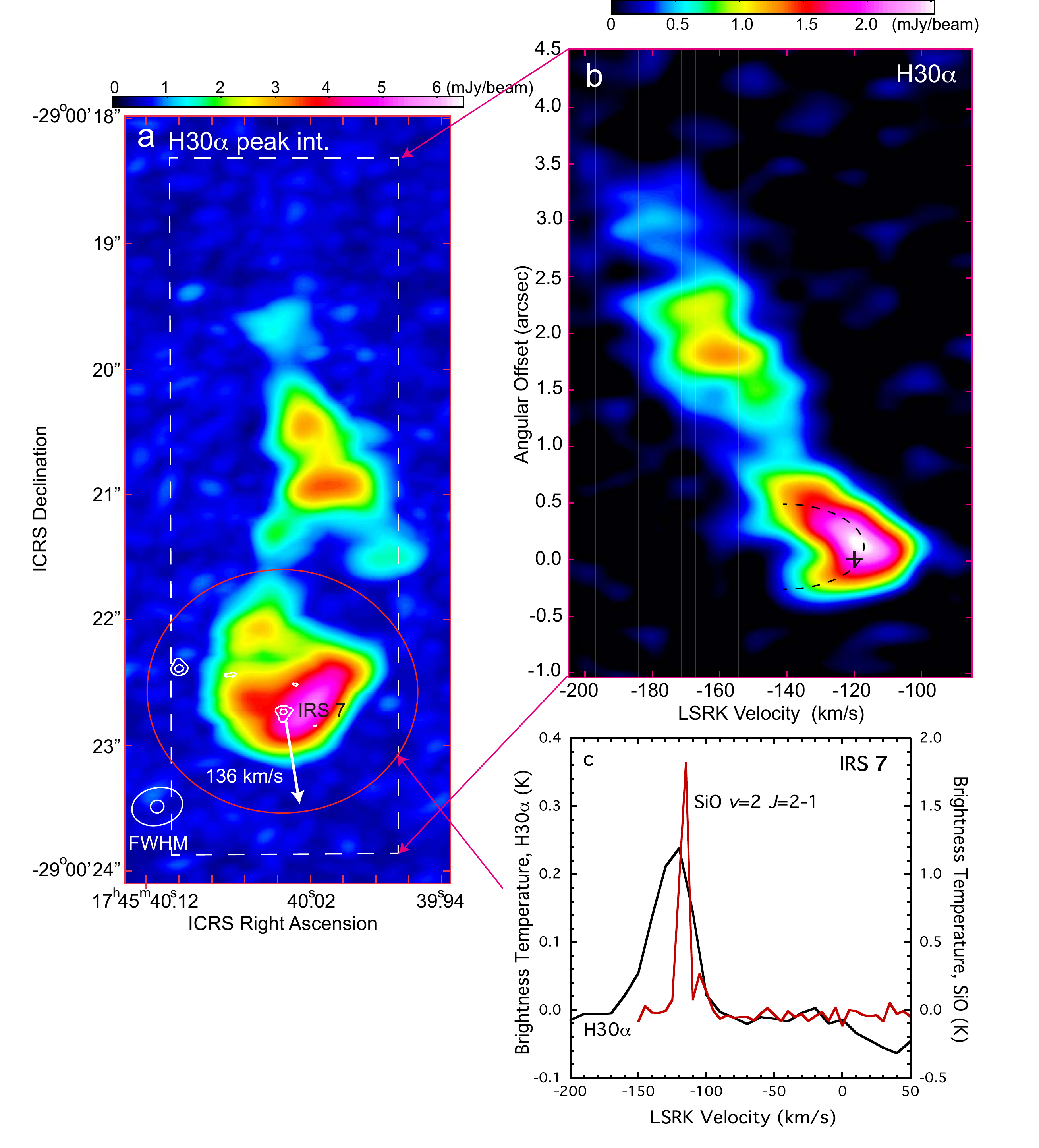}
\end{center}
\caption{{\bf a}  Peak intensity map of IRS 7 and north extension in the H30$\alpha$ recombination line (pseudo color) and 340 GHz continuum emission (contours). The velocity range of the peak intensity map is $V_\mathrm{LSR} = -90$ to $-190$ km s$^{-1}$.  The transverse velocity  of IRS 7, which is shown as a vector, is $V_\mathrm{\perp} = 136$ km s$^{-1}, PA=-170^\circ.6$ based on the proper motion measured by the SiO maser astrometry (\cite{Borkar}).  
{\bf b}  Position-velocity (PV) diagram of IRS 7 in the H30$\alpha$ recombination line along the white broken rectangle in {\bf a}. 
The cross shows the velocity and position of the SiO $v=2, J=2-1$ maser emission toward IRS 7 shown in figure 4c 
 {\bf c} Line profiles toward IRS 7 of the H30$\alpha$ recombination line (black) and  SiO $v=2, J=2-1$ maser emission line (red). The oval in {\bf a} shows the sampling area.
 }
\end{figure}

\section{Discussion}  
\subsection{Photosphere at 340 GHz of IRS 7}
As mentioned in Result, the flux density  at 340 GHz of IRS 7 is $S_\nu=448\pm45 \mu$Jy. 
While the flux density at 34.5 GHz of IRS 7 would be less than $S_\nu<15 \mu$Jy (see Figure 2 in \cite{Yusef-Zadeh2015}).
In these frequencies, absorption is negligible even in the Galactic center region.
The spectrum index is estimated to be $\alpha >1.5 $ if  the spectrum between these frequencies is depicted by a power law, $S_\nu\propto\nu^{\alpha}$.
The large positive value of $\alpha$ suggests that the continuum emission of this frequency range is optically-thick thermal emission. This also suggests that the sub-millimeter continuum emission is emitted from the photosphere itself or the vicinity.

As mentioned in Introduction, IRS 7 would be an M1 or M2 red supergiant star.  
The effective temperature  of the photosphere  is reported to be $T_\mathrm{eff}= 3600\pm200$ K for M1( \cite{Paumard2014}) and $T_\mathrm{eff} = 3600\pm230$ K for M2  (\cite {Carr}).  The brightness temperature at 340 GHz is assumed to be $T_\mathrm{eff} = 3600$ K here. The beam averaged brightness temperature of IRS 7 is measured to be $\bar{T_\mathrm{B}} =0.41\pm0.04$ K as mentioned in Result.  The disagreement between the effective temperature and  brightness temperature is considered to be originated mainly by beam dilution effect which is shown by
\begin{equation}
\label{ }
\bar{T_\mathrm{B}}= \frac{\int\int_\mathrm{IRS 7} T_\mathrm{B}(\theta, \phi)P(\theta, \phi)d\theta d\phi}{\int\int_\mathrm{SB} P(\theta, \phi)d\theta d\phi} , 
\end{equation}
where $P(\theta, \phi)$ and $T_\mathrm{B}(\theta, \phi)$ are the shape function of the synthesized beam and the distribution of the brightness temperature of IRS 7, respectively. Assuming that the distribution of the brightness temperature is like an uniform-disk,  this is approximated as the following equation,
\begin{equation}
\label{ }
\bar{T_\mathrm{B}}\sim T_\mathrm{B}\frac{ \pi(R/d)^2}{1.133\theta_\mathrm{maj} \theta_\mathrm{min}},
\end{equation}
where $R/d$ is the angular radius of the photosphere of IRS 7 at 340 GHz. 
Therefore, the radius of the photosphere is  estimated to $R=1170\pm60 ~R_\odot$ in 2016.  Meanwhile IRS 7 has been reported to be resolved marginally into a uniform-disk with $R=960\pm92 ~R_\odot$ in 2013 in the H band ($\lambda=1.65~\mu\mathrm{m}$) by   the Very Large Telescope Interferometer (VLTI) (\cite{Paumard2014}).  Our estimated photosphere radius at 340 GHz is slightly larger than the radius at H band measured by VLTI. 
However, our radius would be consistent with the VLTI radius because IRS 7 is reported to be variable in size based on the observations in 2008 and 2013 (\cite{Paumard2014}). 

\subsection{Kinematics of the Surrounding Structures of IRS 7}
Figure 4a shows  the peak intensity map of IRS 7 and north extension in the H30$\alpha$ recombination line (pseudo color). The velocity range of the peak intensity map is $V_\mathrm{LSR} = -90$ to $-190$ km s$^{-1}$.   The surrounding structure of IRS 7 in the H30$\alpha$ recombination line is made of the shell like structure and the north extension as mentioned in Result. This is also identified as a shell like structure in the 34.5 GHz continuum map (see Figure 3a in \cite{Yusef-Zadeh2017}). Although this feature has a more sharp edge than that of the H30$\alpha$ recombination line image, the difference of the appearance would be caused mainly by difference of the beam sizes ($\theta_\mathrm{maj} \times \theta_\mathrm{min}= 0\arcsec.09\times 0\arcsec.05$ at 34.5 GHz). However, it cannot completely ruled out another cause, that is the distribution of the electron temperature because the line/continuum ratio is sensitive to the electron temperature.  
In addition, these structures have no associated feature in the molecular emission lines, for example CS $J=2-1$, H$^{13}$CO$^+ J=1-0$, and SiO $v=0, J=2-1$ emission lines (see figures 2, 3, and 4 in \cite{Tsuboi2018}, see also \cite{Moser}, \cite{Yusef-Zadeh2017}).  

Figure 4b shows the position-velocity (PV) diagram of the surrounding structure in the H30$\alpha$ recombination line along the white broken rectangle in Figure 4a (Cf. figure 7b in \cite{Zhao2009}). The shell like structure is also identified  as a half shell-like feature in the diagram (a broken line curve in Figure 4b) . The velocity extent of the structure is as large as $\Delta V_\mathrm{FWZI} \sim 40$ km s$^{-1}$. 
While the north extension is identified  as a linear feature with a velocity gradient of $\sim22$ km s$^{-1}$ arcsec$^{-1}$.

Figure 4c shows the line profile of  the H30$\alpha$ recombination line  toward IRS 7.  A red oval in Figure 4a shows the integration area. 
The shell like structure around IRS 7 has a single-peak line profile. The central velocity and FWHM velocity width by Gaussian fitting is $V_\mathrm{c} = -124.4\pm0.63 $ km s$^{-1}$ and  $V_\mathrm{FWHM} =31.9\pm1.5$ km s$^{-1}$, respectively.  
The ionization of the structure would be made by the huge Lyman continuum photons from the NSC itself (e.g. \cite{Zhao2010}). Assuming Hydrogen ion temperature is equal to the electron temperature, $T^\ast_\mathrm{e}=4650$ K (see the next subsection), the sound velocity of the ionized gas is estimated to $C_\mathrm{s} = 8.9$ km s$^{-1}$ using the formula given by 
\begin{equation}
\label{ }
C_\mathrm{s} =\sqrt{2kT_{\mathrm i}/m_{\mathrm H}} =13\times[T^\ast_\mathrm{e}/10^4\mathrm{K}]^{0.5}
 [\mathrm{km~s}^{-1}].
\end{equation} 
 If the observed velocity extent originates from the gas expansion alone, the expanding velocity is estimated to $V_\mathrm{exp} = 14\pm1$ km s$^{-1}$ using the formula given by
\begin{equation}
\label{ }
 V_\mathrm{exp} =\sqrt{V_\mathrm{FWHM}^2-(2\sqrt{ln2}C_\mathrm{s})^2}/2.
\end{equation} 
The expanding velocity shows that the expansion is moderately supersonic.

Figure 4c also shows the line profile of the SiO $v=2, J=2-1$ maser emission.  This spectrum is extracted from the ALMA data shown in the previous paper (\cite{Tsuboi2017}). There is a peak of the SiO  maser emission line at  $V_\mathrm{LSR} = -120$ km s$^{-1}$. This is consistent with the previous observations  (e.g. \cite{Reid2003},  \cite{Borkar}), and is also consistent with the  central velocity of the ionized gas (a cross in Figure 4b).
The SiO maser emissions from late-type stars are thought to be emitted from the atmospheres in the vicinity of the stellar surfaces.  The radial velocity and position of the SiO maser emission should be nearly equal to the radial velocity and position of IRS 7 itself.  The velocity is centered on that of the shell like structure, while the position is slightly shifted to the south.
While the transverse velocity  of IRS 7 is reported to be $V_\perp = 136$ km s$^{-1}, PA=-170^\circ.6$ based on the proper motion measured by the SiO maser astrometry (\cite{Borkar}, see also \cite{Reid2003}).  The transverse velocity  is shown as a vector in figure 4a. 
Although the half bright portion of the shell like structure is slightly inclined (see figure 4a), the vector is seen to be perpendicular to it. 
The spatial traveling velocity of IRS 7  is estimated to be $V_\mathrm{t}\sim181$ km s$^{-1}$ using the formula $V_\mathrm{t}=\sqrt{V_\perp^2+V_\mathrm{LSR}^2}$. 


The north extension would be a gas stream flowed from the shell like structure, or would be like a cometary tail. 
This is because both the north extension and the shell like structure are included in the inclined linear feature in the PV diagram. 
The velocity of the southernmost part of the north extension  is nearly equal to the northernmost part of the shell like structure. 
The velocity gradually decreases with increasing angular offset and becomes equal to that of the ambient matter, $V_\mathrm{LSR} \sim -170 $ km s$^{-1}$, finally.
The radial velocity difference between IRS 7 and the ambient matter, or the radial airspeed, is guessed to be $V_\mathrm{air, \parallel} \sim 50$ km s$^{-1}$. 
The 3D airspeed would be at least $V_\mathrm{air} \gtrsim 50$ km s$^{-1}$ or supersonic. 
Note that  this is considerably larger than the derived expanding velocity of the ionized gas, $V_\mathrm{air}>V_\mathrm{exp} = 14$ km s$^{-1}$. 

IRS 7 releases gas as stellar wind by pulsation (\cite{Paumard2014}). The shell-like structure is made by the stellar wind from IRS 7, which is ionized by the strong FUV from the NSC. IRS 7 runs in the ambient matter at a supersonic speed in the south direction.  The ionized gas is left behind and looks like the cometary tail mentioned above. Although CO absorption lines are observed in the atmosphere of IRS7 (e.g. \cite{Paumard2014}), the surrounding structure is associated with no feature in the molecular emission lines as mentioned in Results. 
This would be caused by that the strong FUV from the NSC also destroyed such molecules in the stellar wind.
 
The transverse velocity vector is seen to be perpendicular to the brighten southern half.
This relation suggests that the brightened southern half of the structure is a sign of the bow shock made by supersonic gas traveling in the ambient matter.  The positional shift of IRS 7 to the south in the shell like structure (see Result)  is consistent with the bow shock hypothesis, which  have already been discussed (\cite{Serabyn},  \cite{Yusef-Zadeh1991}, \cite{Yusef-Zadeh1992}). 
The kinetic energy of a Hydrogen atom traveling to the south in the bow shock is simply estimated using the following equation;
 \begin{equation}
\label{ }
 E_\mathrm{kin} =\frac{1}{2}m_\mathrm{H}(V_\mathrm{air}+V_\mathrm{exp})^2.
\end{equation} 
The kinetic energy reaches to $E_\mathrm{kin} \sim 22~\mathrm{eV}$, which is larger than the ionization energy of a Hydrogen atom of $E_\mathrm{H, ion}=13.6~\mathrm{eV}$.  Therefore such shock wave could induce the collisional ionization of the southern half in addition to the ionization by FUV from the NSC. 

\subsection{Physical Properties of the Surrounding Structures of IRS7}
The LTE electron temperature, $T^\ast_{\mathrm e}$, of the shell like structure around IRS 7 is estimated from the ratio between the integrated recombination line intensity, $\int S_{\mathrm{line}}(\mathrm{H}30\alpha)dv$, and the continuum flux density, $S_\nu(\mathrm{232GHz})$, assuming that the line and continuum emissions are both optically thin and in local thermodynamic equilibrium. 
The integrated  intensity of the recombination line is derived to be  $\int_{-95}^{-155} S_{\mathrm{line}}(\mathrm{H}30\alpha)dv =0.79\pm0.08$ Jy km s$^{-1}$.  The continuum flux density is estimated to be $S_\nu(\mathrm{232GHz})=3.81\pm0.10$ mJy using $S_\nu(\mathrm{232GHz})=S_\nu(\mathrm{340GHz})(\frac{340}{232})^{0.1}$.
The integration area of these values is shown in Figure 4a as an oval. As shown in figure 4c, the shell like structure has a single-peak line profile,  thus it is easy to derive the electron temperature of the structure because it is not contaminated with other velocity components.
The well-known formula of the LTE electron temperature is given by
\begin{equation}
\label{1}
T^\ast_{\mathrm e}[\mathrm K]=\left[\frac{6.985\times10^3}{a(\nu, T^\ast_{\mathrm e})}\Big(\frac{\nu}{\mathrm{GHz}}\Big)^{1.1}
\frac{1}{1+\frac{N(\mathrm{He^+})}{N(\mathrm{H^+})}}
\frac{S_\nu(\nu)}
{\int S_{\mathrm{line}}\Big(\frac{dv}{\mathrm{km~s}^{-1}}\Big)}
\right]^{\frac{1}{1.15}}.
\end{equation}
The correction factor, $a(\nu, T^\ast_{\mathrm e})$, at $\nu=232$ GHz is calculated to be $a=0.678-0.892$  for $T^\ast_{\mathrm e}=0.3-1.5\times10^4$ K \citep{Mezger}.  We assume that the number ratio of He$^+$ to H$^+$ is $\frac{N(\mathrm{He^+})}{N(\mathrm{H^+})}=0.09$, a typical value in the Galactic center region (e.g. \cite{Tsuboi2017}). The LTE electron temperature is obtained by iteratively solving the formula for $T^\ast_{\mathrm e}$. 
The mean electron temperature of the shell like structure around IRS 7 is estimated to $\bar{T}^\ast_{\mathrm e}=4650\pm500$ K.
The estimated temperature is  fairly lower than those of the previous observations (e.g. $\bar{T}^\ast_{\mathrm e}=7000$ K in \cite{Zhao2010}). This is slightly higher than the effective temperature  of the photosphere of IRS 7, $T_\mathrm{eff}=3600$ K. 

The mean brightness temperature of the continuum emission is estimated to be  $\bar{T}_{\mathrm B}=8.0\pm0.1$ K using $\bar{T}_{\mathrm B}=1.22\times10^6\Big(\frac{\theta_{\mathrm d}}{\mathrm{arcsec}} \Big)^{-2}\Big(\frac{\nu}{\mathrm{GHz}}\Big)^{-2}S_\nu$. The electron density, $\bar{n}_{\mathrm e}$.  of the ionized gas ring around IRS 7 is estimated by the following well-known formula;
\begin{equation}
\label{2}
\bar{n}_{\mathrm e}[{\mathrm cm}^{-3}]=\left[\frac{\bar{T}_{\mathrm B}T_{\mathrm e}^{\ast 0.35}\Big(\frac{\nu}{\mathrm{GHz}}\Big)^{2.1}}{2.674\times10^{-20}\alpha(\nu,T)\Big(\frac{L}{\mathrm{cm}}\Big)}\right]^{0.5}
\end{equation}
\citep{Altenhoff}.
The mean path length is given by $L=\frac{4\pi}{3}(\frac{D}{2})^3/\pi(\frac{D}{2})^2=\frac{2}{3}D$ assuming that the ionized gas has a spherical shape with the diameter, $D$. The diameter is $D\sim0.055$ pc$=1.7\times10^{17}$ cm (see  Result). The mean path length is $ L\sim1.1\times10^{17}$ cm. The mean electron density is estimated to be $\bar{n}_{\mathrm e}=(6.1\pm0.6)\times10^4$ cm$^{-3}$.
Therefore the mass of the released gas from IRS 7 is estimated using the formula;
\begin{equation}
\label{2}
 M_{\mathrm gas}=\frac{4\pi}{3}\Bigg(\frac{D}{2}\Bigg)^3\bar{n}_{\mathrm e}m_{\mathrm{H}}\frac{X(\mathrm{H})+X(\mathrm{He})}{X(\mathrm{H})},
 \end{equation}
where $m_{\mathrm{H}}$ and $\frac{X(\mathrm{H})+X(\mathrm{He})}{X(\mathrm{H})}$ are the mass of a Hydrogen atom and the mass abundance ratio, respectively.  The mass abundance ratio is $\frac{X(\mathrm{H})+X(\mathrm{He})}{X(\mathrm{H})}\sim1.4$ assuming the solar abundance of elements. The mass of the ionized gas is $M_{\mathrm gas}\sim0.18 M_\odot$
As mentioned in the previous subsection, the expanding velocity of the ionized gas is  $V_\mathrm{exp} = 14$ km s$^{-1}$. The time scale of the expansion is estimated to be $T_\mathrm{exp} \sim \frac{D/2}{V_\mathrm{exp}} =1900$ yrs. If the gas mass is  maintained mainly by the mass loss of IRS 7, the mass loss rate is estimated to be $\dot{m} \sim  \frac{M_{\mathrm gas}}{T_\mathrm{exp}}=1\times 10^{-4} M_\odot$ yr$^{-1}$. This value is consistent with the expected mass loss rate of a pulsating red supergiant star with $M=20-25 M_\odot$ (e.g. \cite{Yoon}).

\begin{ack} 
This work is supported in part by the Grant-in-Aids from the Ministry of Eduction, Sports, Science and Technology (MEXT) of Japan, No.19K03939 (MT and MM).
The National Radio Astronomy Observatory (NRAO) is a  facility of the National Science Foundation operated under cooperative  agreement by Associated Universities, Inc. This paper also makes use of the following ALMA data:ADS/JAO.ALMAA\#2015.1. 01080.S and ALMA\#2015.A.00021.S.    ALMA is a partnership of ESO (representing its member states), NSF (USA) and NINS (Japan), together with NRC(Canada), NSC and ASIAA (Taiwan), and KASI (Republic of Korea), in cooperation with the Republic of Chile. The Joint ALMA Observatory is operated by ESO, AUI/NRAO and NAOJ.  
\end{ack}

\clearpage


\begin{thebibliography}{}
\bibitem[Altenhoff  et~al.(1960)]{Altenhoff}Altenhoff, W.,  et~al. \ 1960, Ver\"off. d. Sternw. Bonn Nr.59
\bibitem[Boehle et~al. (2016)] {Boehle}Boehle, A. et~al.\ 2016, \apj, 830, 17
\bibitem[Borkar et~al. (2019)] {Borkar}Borkar, A., Eckart, A., Straubmeier, C., Sabha, N.B., et~al.
\ 2019, arXiv:1909.13753
\bibitem[Carr, Sellgren, \& Balachandran (2000)] {Carr}Carr, J.S., Sellgren, K., \& Balachandran, S.C., \  2000,  \apj, 530, 307
\bibitem[Figer et~al. (1999)]{Figer1999} Figer, D.~F.; McLean, I.~ S., \& Morris, M., \ 1999, \apj, 514, 202
\bibitem[Figer et~al. (2002)]{Figer2002} Figer, D.~F.  Najarro, F., Gilmore, D., Morris, et~al.
\ 2002, \apj, 581, 258
\bibitem[Genzel et~al. (1996)]{Genzel}Genzel, R., Thatte, N., Krabbe, A., Kroker, H., \& Tacconi-Garman, L. E. \ 1996, \apj, 472, 153
\bibitem[Ghez et~al. (2008)]{Ghez}Ghez, A. M., et al. \ 2008, \apj, 689, 1044
\bibitem[Gravity Collaboration (2018)] {Abuter}Gravity Collaboration, \ 2018, \aap, 615, L15
\bibitem[Kepley  et~al. (2019)]{Kepley}Kepley, A. A., Tsutsumi, T., Brogan, C. L., Indebetouw, R., Yoon, I., Mason, B. and Meyer, J. D.  2019, arXiv:1912.04970
\bibitem[McMullin et~al. (2007)]{McMullin}McMullin, J. P., Waters, B., Schiebel, D., Young, W., \& Golap, K. \ 2007, Astronomical Data Analysis Software and Systems XVI (ASP Conf. Ser. 376), ed. R. A. Shaw, F. Hill, \& D. J. Bell (San Francisco, CA: ASP), 127 
\bibitem[Mezger \& Henderson (1967)]{Mezger}Mezger, P. G.\& Henderson, A. P. \ 1967, \apj, 147, 471 
\bibitem[Moser et~al.  (2017)]{Moser}Moser, L., S\'anchez-Monge, \'A., Eckart, A., Requena-Torres, M. A., Garc\'{\i}a-Marin, M., Kunneriath, D., Zensus, A., Britzen, S., Sabha, N., Shahzamanian, B., Borkar, A., \& Fischer, S., \ 2017, \aap, 603, A68
\bibitem[Paumard et~al. (2014)]{Paumard2014} Paumard, T. ,  Pfuhl, O.,  Martins, F.,   Kervella, P., Ott, T.,  Pott, J.-U. et al.,
\ 2014, \aap, 568, A85
\bibitem[Perger et~al. (2014)]{Perger}Perger, M.,   Moultaka, J.,   Eckart, A.,  Viehmann, T.,  Sch\"odel, R. and K. Muzic,  \ 2008, \aap, 478, 127
\bibitem[Reid et~al. (2003)]{Reid2003}Reid, M. J., Menten, K. M., Genzel, R., Ott, T., Sch\"odel, R., \& Eckart, A., \ 2003, \apj, 587, 208
\bibitem[Sch\"odel, Merritt \& Eckert (2009)]{Schodel} Sch\"odel, R., Merritt, D., \& Eckart, A., \ 2009, \aap, 502, 91
\bibitem[Serabyn, Lacy and Achtermann (1991)]{Serabyn}Serabyn, E., Lacy, J. H., \& Achtermann, J. M. 1991, \apj, 378, 557
 \bibitem[Tsuboi et~al. (2017)]{Tsuboi2017}Tsuboi, M., Kitamura, Y., Uehara, K., Miyawaki, R., Tsutsumi, T., Miyazaki, A.,  \& Miyoshi, M.,    \ 2017, \apj,  842, id. 94
\bibitem[Tsuboi et~al. (2017b)]{Tsuboi2017b}Tsuboi, M., Kitamura, Y., Tsutsumi, T., Uehara, K., Miyoshi, M., Miyawaki, R., \& Miyazaki, A., \ 2017b,  \apjl, 850, id L5
\bibitem[Tsuboi et~al. (2018)]{Tsuboi2018}Tsuboi, M., Kitamura, Y., Uehara, K., Tsutsumi, T.,   Miyawaki, R., Miyoshi, M., \& Miyazaki, A., \ 2018,  \pasj, 70, id 85
\bibitem[Yoon \& Cantiello (2010)]{Yoon}Yoon, S-C. and  Cantiello, M.,  \ 2010, \apjl, 717, L62
\bibitem[Yusef-Zadeh  \& Morris (1991)]{Yusef-Zadeh1991}Yusef-Zadeh, F.,  \& Morris, M. 1991, \apjl, 371, L59
\bibitem[Yusef-Zadeh  \& Melia (1992)]{Yusef-Zadeh1992}Yusef-Zadeh, F. \& Melia, F., \ 1992, \apjl, 385, L41
\bibitem[Yusef-Zadeh et~al. (2014)]{Yusef-Zadeh2014}Yusef-Zadeh, F., Roberts, D. A.,Bushouse, H., Wardle, M., Cotton, W., Royster, M.,\& van Moorsel, G., \ 2014, \apjl, 792, id. L1
\bibitem[Yusef-Zadeh et~al. (2015)]{Yusef-Zadeh2015}Yusef-Zadeh, F.,  Bushouse, H.,  Sch\"odel, R.,  Wardle, M.,  Cotton, W.,  Roberts, D. A., Nogueras-Lara, F. \& Gallego-Cano, E. \ 2015, \apj, 809, id 10.
\bibitem[Yusef-Zadeh et~al. (2017)]{Yusef-Zadeh2017}Yusef-Zadeh, F.,  Wardle, M.,  Kunneriath, D.,  Royster, M.,  Wootten, A., \&. \ 2017, \apjl, 850, id. L30
\bibitem[Zhao et~al. (2009)]{Zhao2009}Zhao, J.-H., Morris, M. R., Goss, W. M., An, T., \ 2009, \apj, 699,186
\bibitem[Zhao et~al. (2010)]{Zhao2010}Zhao, J.-H., Blundell, R., Moran, J. M., Downes, D., Schuster, K. F., Marrone, D. P., \ 2010, \apj, 723,1097
\end{thebibliography}
\end{document}